\documentstyle[12pt]{article}
   \font\tenmsb=msbm10 scaled\magstep 1
   \font\sevenmsb=msbm7 scaled \magstep 1
   \font\faivemsb=msbm5 scaled \magstep 1
\newfam\msbfam
      \textfont\msbfam=\tenmsb
      \scriptfont\msbfam=\sevenmsb
      \scriptscriptfont\msbfam=\faivemsb
\def\Bbb#1{{\fam\msbfam #1}}
\font\tengothic=eufm10 scaled\magstep 1
\font\sevengothic=eufm7 scaled\magstep 1
\newfam\gothicfam
      \textfont\gothicfam=\tengothic
      \scriptfont\gothicfam=\sevengothic

\newcommand{\be}{\begin{equation}}
\newcommand{\ee}{\end{equation}}
\newcommand{\prt}{\partial}
\newcommand{\dlt}{\delta}
\newcommand{\lbd}{\lambda}
\newcommand{\vp}{\varphi}
\newcommand{\ra}{\rightarrow}
\newcommand{\Dlt}{\Delta}
\newcommand{\ga}{\gamma}
\newcommand{\om}{\omega}
\newcommand{\sr}{\stackrel{\rightarrow}{r}}
\newcommand{\dgr}{\dagger}
\newcommand{\lgl}{\langle}
\newcommand{\rgl}{\rangle}
\newcommand{\ep}{\varepsilon}

\begin{document}

\begin{center}
{\large{\bf Local Stability of Dynamical Processes in Random Media} \\ [5mm]
V.I.Yukalov} \\ [3mm]
{\it Department of Mathematics and Statistics \\
Queen's University, Kingston, Ontario K7L 3N6, Canada \\
and \\
Bogolubov Laboratory of Theoretical Physics \\
Joint Institute for Nuclear Research, Dubna, Russia}
\end{center}

\vspace{8cm}

Short title: Local stability of dynamical processes

\vspace{1cm}

{\bf PACS:} 02.50.Ey, 05.40.+j, 02.30.Jr

\vspace{5mm}

{\bf Keywords:} Stochastic processes, Local stability, 

\hspace{2.2cm} Partial differential equations

\newpage

\begin{abstract}

A particular type of random dynamical processes is considered, in which the stochasticity is introduced through randomly fluctuating parameters.
A method of local multipliers is developed for treating the local stability 
of such dynamical processes corresponding to infinite--dimensional dynamical 
systems. The method is illustrated by several 
examples, by the random diffusion equation, random wave equation, 
and random Schr\"odinger equation. The evolution equation for the 
density matrix of a quasiopen statistical system subjest to the action of
random surrounding is considered. The stationary solutions to this 
equation are found to be unstable against arbitrary small finite 
random perturbations. The notion of random structural stability is 
introduced.

\end{abstract}

\newpage

\section{Introduction}

The stability of dynamical systems described by ordinary differential 
equations can be characterized by Lyapunov exponents. There are several kinds 
of the latter, for example, the classical Lyapunov exponents characterizing asymptotic stability [1,2], the finite--interval exponents describing 
stability on finite intervals [3,4], the pointwise local--time Lyapunov
exponents giving information on the local stretching and contraction rates 
of trajectories [5,6], and generalized Lyapunov exponents taking into 
account correlations [7].

More difficult is the situation with the so--called infinite--dimensional 
dynamical systems modelled by partial differential equations. There is 
no general stability theory for such equations, although 
in many cases one can analyse the asymptotic stability by linearizing 
these equations about a stationary solution [8].

Even more complicated is the case when partial differential equations 
contain random fields. The present paper addresses just this type 
of equations describing random dynamical processes. More precisely, only one particular type of these processes will be considered here, when the stochasticity is introduced into differential equations by means of randomly distributed parameters. Such a kind of equations may be called {\it parametrically random differential equations}. This is a specific case of stochastic differential equations. The latter are usually treated as containing Gaussian stochastic fields related to the Wiener or Ornstein--Uhlenbeck processes [9-12]. These will not be touched on in the paper --  we shall deal only with parametrically random equations. And, although the words random and stochastic mean the same, we shall rather use the former instead of the latter in order to distinguish our case from what one usually implies under stochastic processes. 

Parametrically random equations form an important class of equations that can be often met in applications, such as quantum electronics. Thus, the influence of interference on propagating signals may be modelled by randomly fluctuating amplitude and phase. The random shift of frequency describes various local defects leading to the so--called nonuniform broadening. Some local interactions can be treated as random forces whose existence may become incomparably more essential for a nonequilibrium system than the presence of the Nyquist noise. For example, such random local fields, due to nonsecular dipole interactions, trigger pure spin superradiance in nonequilibrium nuclear magnets [13,14].

The approach suggested below may be generalized in several aspects. However, for the sake of clarity, here I prefer to stick myself to the parametrically random dynamical processes. To treat the stability of the latter, the method 
of multipliers is developed. For ordinary differential equations without 
random fields the method of multipliers is completely
equivalent to that of the Lyapunov exponents. However, for partial 
differential equations, especially with random fields, the method 
of multipliers is more natural and general: one may define multipliers 
when it is difficult or even impossible to introduce Lyapunov exponents.
Multipliers characterize both the local as well as asymptotic stability. 
They can describe not only the exponential stability, as the Lyapunov 
exponents do, but also other types of stability. The effectiveness of 
using multipliers for analysing the local stability of dynamical processes 
has been demonstrated earlier for approximation cascades and approximation 
flows [15-17].

\section{Method of local multipliers}

Let $\;x\in\Bbb{D}\;$ be a set of continuous variables pertaining to a 
domain $\;\Bbb{D}\;$ which can be bounded or not; let $\;t\in\Bbb{R}_+\;$ 
denote time; and let $\;\xi\;$ be a set of random variables with a probability
distribution $\;p(\xi)\;$. Consider a set, enumerated by the index
$\;i=1,2,\ldots,d\;$, of real or complex functions $\;y_i(x,\xi,t)\;$ of 
these variables. Define the column
$$ y(\xi,t)=[y_i(x,\xi,t)] $$
with respect to $\;i\;$ and $\;x\;$. This column is assumed to be a solution 
of the evolution equation
\be
\frac{\prt}{\prt t}y(\xi,t) =v(y,\xi,t)
\ee
with the velocity field
$$ v(y,\xi,t)=[v_i(x,y,\xi,t)] $$
which may contain differential and integral operators acting on 
$\;y(\xi,t)\;$. The evolution equation (1) is to be supplemented by the 
initial condition
\be
y(\xi,0) = f = [f_i(x)] ,
\ee
in which $\;f_i(x)\;$ are given functions, and by boundary conditions. 
The quantity of final interest is the solution
\be
y(t)\equiv\int y(\xi,t)p(\xi)d\xi 
\ee
averaged over random variables.

Equation (1) is a compact form of writing down a wide class of evolution
equations. In particular, it includes the finite $\;d\;$--dimensional 
dynamical systems described by ordinary differential equations. But in 
general, this class embraces infinite--dimensional dynamical systems 
with various partial differential equations, such as can be met in 
considering hydrodynamic and plasma turbulence [18,19], dynamics of structures
in liquid flows [20], soliton dynamics in condensed matter [21], 
electromagnetic radiation in large systems [22-24], and phase--ordering 
kinetics in quenched systems [25].

The stability of motion is related to the variation of the solution $\;y(\xi,t)\;$ at time $\;t\;$,
\be
\dlt y(\xi,t) =M(\xi,t,t_0)\dlt y(\xi,t_0) ,
\ee
with respect to its variation at time $\;t_0\;$. Here the {\it multiplier matrix}
$$ M(\xi,t,t_0) = [M_{ij}(x,x',\xi,t,t_0)] $$
is introduced with the elements
\be
M_{ij}(x,x',\xi,t,t_0) \equiv\frac{\dlt y_i(x,\xi,t)}{\dlt y_j(x',\xi,t_0)} .
\ee
From this definition it follows that 
$$ M_{ij}(x,x',\xi,t,t) =\dlt_{ij}\dlt(x-x') . $$
Therefore, at the coinciding times, $\;t_0=t\;$, the multiplier matrix is 
the unity matrix
\be
M(\xi,t,t) =\hat 1 \equiv [\dlt_{ij}\dlt(x-x')]
\ee
in the space of the indices $\;i\;$ and variables $\;x\;$. Remind that in (4)
the matrix notation is used according to which the right--hand side of (4) 
is the column
$$ M(\xi,t,t_0)\dlt y(\xi,t) = \left [\sum_{k}\int M_{ik}(x,x',\xi,t,t_0)
\dlt y_k(x',\xi,t_0)dx'\right ] . $$ 
From the variational--derivative property
$$ \frac{\dlt y_i(x,\xi,t)}{\dlt y_j(x',\xi,t_0)} =\sum_{k}\int
\frac{\dlt y_i(x,\xi,t)}{\dlt y_k(x_1,\xi,t_1)}
\frac{\dlt y_k(x_1,\xi,t_1)}{\dlt y_j(x',\xi,t_0)}dx_1 $$
we find that
\be
M(\xi,t,t_1) M(\xi,t_1,t_0) = M(\xi,t,t_0) .
\ee
Putting in (7) $\;t_0=t\;$ and using (6), we obtain the definition of 
the inverse multiplier matrix
\be
M^{-1}(\xi,t,t_0) =M(\xi,t_0,t) .
\ee
The properties (6)--(8) show that the multiplier matrices form a group
$$ {\cal M}(\xi) =\left\{ M(\xi,t,t_0)\;|t,t_0\in\Bbb{R}_+\right\} , $$
which may be called the {\it multiplier group}. It is also evident that 
the multiplier matrices are the evolution operators for the variation 
$\;\dlt y(\xi,t)\;$.

If the random variable $\;\xi\;$ were fixed, then the transformation (4)
would be contracting provided that $\;||M(\xi,t,t_0)|| < 1\;$. However, 
this does not yield, in general, the stability of the averaged solution (3).

Variating the evolution equation (1), we get the equation
\be
\frac{\prt}{\prt t}M(\xi,t,t_0) = L(y,\xi,t)M(\xi,t,t_0)
\ee
for the multiplier matrix, where the matrix
$$ L(y,\xi,t) = [L_{ij}(x,x',y,\xi,t)] $$
consists of elements
\be
L_{ij}(x,x',y,\xi,t) \equiv
\frac{\dlt v_i(x,y,\xi,t)}{\dlt y_j(x',\xi,t)} . 
\ee
In particular cases, the real parts of the eigenvalues of the matrix
$\;L(y,\xi,t)\;$ define the Lyapunov exponents, because of which we shall 
call it the {\it Lyapunov matrix}. The property (6) plays the role of 
the initial condition for eq.(9). The boundary conditions for eq. (9) are 
to be obtained by variating those of eq.(1). For example, if the functions
$\;y_i(x,\xi,t)\;$ are given at the boundary of $\;\Bbb{D}\;$, denoted 
by $\;\prt\:\Bbb{D}\;$, then
$$ M_{ij}(x,x',\xi,t,t_0) = 0 \qquad (x\in\prt\:\Bbb{D}) , $$
and if $\;\prt y_i/\prt x\;$ is given at $\;\prt\:\Bbb{D}\;$, then
$$ \frac{\prt}{\prt x}M_{ij}(x,x',\xi,t,t_0) = 0 \qquad (x\in\prt\:\Bbb{D}). $$

Generally, averaging the variation (4) over the random variables does not
lead a simple relation between the variation of (3) and an averaged 
multiplier matrix, except for the case $\;t_0=0\;$, when $\;y(\xi,0)\;$ is 
fixed by the initial condition (2) not containing $\;\xi\;$. In such a case 
the variation of (3) is
\be
\dlt y(t) = M(t)\dlt f
\ee
with the averaged multiplier matrix
\be
M(t) \equiv \int M(\xi,t,0)p(\xi)d\xi .
\ee
The transformation (11) is contracting provided that $\;||M(t)||<1\;$, 
from where $\;||\dlt y(t)||<||\dlt f||\;$. Then we may say that the motion 
is locally stable on the time interval $\;[0,t]\;$.

As we see, to characterize the stability of motion we need to know 
the multiplier matrix. If we are interested only in sufficient conditions 
of stability, we can evaluate the norm $\;||M(t)||\;$ by employing various
inequalities [26] valid for the solutions of the matrix equations of the 
type (9). Since the multiplier matrix satisfies eq.(9) involving the 
Lyapunov matrix, the properties of these matrices are closely interrelated. 
Some properties of the multiplier matrices are considered in Appendix.
Throughout the paper we shall mainly deal with the case when the Lyapunov 
matrix meets the following two conditions greatly simplifying the analysis.

\vspace{3mm}

{\bf Condition 1.} The Lyapunov matrix is stationary:
\be
L(y,\xi,t) = L(\xi) \equiv [L_{ij}(x,x',\xi)] ,
\ee
that is, does not depend on time.

\vspace{3mm}

{\bf Condition 2.} The Lyapunov matrix possesses a set of eigenfunctions forming 
a complete stochastically invariant basis:
\be
L(\xi)\vp_n =\lbd_n(\xi)\vp_n , \qquad \vp_n =[\vp_{ni}(x)] ,
\ee
which means that the eigenvalue problem (14) has solutions with 
the eigenfunctions forming a complete basis independent of the random
variable $\;\xi\;$.

If these conditions are not fulfilled for eq.(9), it may happen, nevertheless,
that there exists a transformation reducing (9) to an equivalent equation 
with an effective Lyapunov matrix enjoying conditions (13) and (14). 
Another important case is when these conditions are fulfilled approximately,
thus, defining an initial approximation for perturbation theory.

Assuming condition 1, one immediately gets the solution
\be
M(\xi,t,t_0)=\exp \{ L(\xi)(t-t_0)\}
\ee
to the matrix equation (9). From here, under condition (2), it follows that 
the multiplier and Lyapunov matrices possess a common set of eigenfunctions,
\be
M(\xi,t,t_0)\vp_n =\mu_n(\xi,t,t_0)\vp_n ,
\ee
with their eigenvalues related as
\be
\mu_n(\xi,t,t_0) =\exp\{\lbd_n(\xi)(t-t_0)\} .
\ee
The eigenvalues of the multiplier matrix will be called the {\it local multipliers}. As is obvious from (17),
\be
\mu_n(\xi,t,t) = 1.
\ee

With a complete basis $\;\{\vp_n\}\;$, we may define the expansions for 
the solution
\be
y(\xi,t) =\sum_{n}c_n(\xi,t)\vp_n
\ee
and for its initial value
\be
f = \sum_{n}f_n\vp_n ,
\ee
where, according to the initial condition (2),
$$ c_n(\xi,0)=f_n . $$
The coefficient functions $\;c_n(\xi,t)\;$ in the expansion (19) have 
the meaning of natural variables, or {\it natural components}, of the 
solution $\;y(\xi,t)\;$. For these components, from eq.(4) we find 
the variation
\be
\dlt c_n(\xi,t) =\mu_n(\xi,t,t_0)\dlt c_n(\xi,t_0) .
\ee

The expansion for the averaged solution (3) reads
\be
y(t) =\sum_{n}c_n(t)\vp_n
\ee
with the averaged components
\be
c_n(t)\equiv \int c_n(\xi,t)p(\xi)d\xi .
\ee
Averaging eq.(16), we come to the  eigenproblem
\be
M(t)\vp_n =\mu_n(t)\vp_n
\ee
showing that the averaged multiplier matrix (12) has the eigenvalues
\be
\mu_n(t) =\int\mu_n(\xi,t,0)p(\xi)d\xi .
\ee
According to (18), we have $\;\mu_n(0) = 1\;$.

Using (22)--(25), we come to the conclusion that the variation (11) is equivalent to the set of variations
\be
\dlt c_n(t) =\mu_n(t)\dlt f_n
\ee
for the natural components (23). The apparent form of (26), in which 
the averaged multiplier (25) is just a function, makes it possible to 
classify the local stability properties at time $\;t\;$.

We shall say that the $\;n\;$--component at time $\;t\;$ is locally stable,
locally neutral, or locally unstable against the variation of initial 
conditions if, respectively,
$$ |\mu_n(t)|<1 \qquad (locally\; stable) , $$
\be
|\mu_n(t)| = 1 \qquad (locally\; neutral) ,
\ee
$$ |\mu_n(t)|>1 \qquad (locally\; unstable) . $$
The motion as a whole, or the process, will be called locally stable, 
locally neutral or locally unstable according to whether
$\;\sup_{n}|\mu_n(t)|\;$ is less than one, equal to or more than one.

If the property of local stability, neutrality or instability of 
an $\;n\;$--component holds at each point $\;t\;$ of an interval
$\;[t_1,t_2]\;$, then we say that the $\;n\;$--component is, respectively, 
{\it uniformly stable, uniformly neutral} or {\it uniformly unstable} on 
this interval. Similarly, the process is uniformly stable, uniformly neutral 
or uniformly unstable on an interval $\;[t_1,t_2]\;$, if the corresponding
property for $\;\sup_n|\mu_n(t)|\;$ holds at each point $\;t\;$ of 
that interval.

In the case of local neutrality of an $\;n\;$--component the latter may be 
found to be unstable with respect to higher--order multipliers defined 
as second, third or higher variational derivatives.

As time $\;t\ra\infty\;$, the $\;n\;$--component can be asymptotically 
stable, neutral or unstable if, accordingly,
$$ \mu_n(t)\ra 0 \qquad (asymptotically\; stable) , $$
\be
|\mu_n(t)| \in (0,\infty) \qquad (asymptotically\; neutral) , 
\ee
$$ |\mu_n(t)|\ra \infty \qquad (asymptotically\; unstable) . $$
The process is asymptotically stable, asymptotically neutral or 
asymptotically unstable if $\;\sup_n|\mu_n(t)|\;$ tends to zero, remains 
finite or tends to infinity as $\;t\ra\infty\;$. Note that
$\;\lim_{t\ra\infty}|\mu_n(t)|\;$ may not exist in the case of 
asymptotic neutrality.

Asymptotic stability can be of different types depending on the law by 
which $\;\mu_n(t)\;$ tends to zero as $\;t\ra\infty\;$. For instance, this 
can be exponential decrease, power--law decay or another tendency. 
Recall the possibility of the polynomial asymptotic stability [27] of solutions to stochastic differential equations. Therefore, the multipliers allow a 
more refined description of stability than the Lyapunov exponents, since 
the latter describe only the exponential asymptotic stability.

Moreover, in the case of random processes the averaged multiplier (25) is,
actually, the sole natural characteristic for defining stability. 
This follows from the fact that the stability properties (27) and (28) 
are directly related to the multiplier $\;\mu_n(t)\;$, but not to 
the eigenvalues of the Lyapunov matrix $\;\lbd_n(\xi)\;$, these two 
quantities being connected through the integral
\be
\mu_n(t) =\int\exp\{\lbd_n(\xi)t\}p(\xi)d\xi .
\ee

One can, of course, introduce effective Lyapunov exponents, such as 
effective finite--interval exponents
$$ \lbd_n^{eff}(t,t_0)\equiv \frac{1}{t-t_0}\ln |\mu_n(t,t_0)| , $$
where
$$ \mu_n(t,t_0)\equiv \int\mu_n(\xi,t,t_0)p(\xi)d\xi , $$
effective pointwise exponents
$$ \lbd_n^{eff}(t,t) \equiv 
\lim_{\Dlt t\ra 0}\frac{1}{\Dlt t}\ln|\mu_n(t+\Dlt t,t)| $$
and effective asymptotic exponents
$$ \lbd_n^{eff}\equiv\lim_{t\ra\infty}\lbd_n^{eff}(t,t_0) . $$
However, these would be excessive unnecessary definitions.

What useful we can get from defining the effective Lyapunov exponents is the
following. Assume that $\;\{\lbd_n\}\;$ is a set of Lyapunov exponents 
ordered in the nonincreasing law, that is $\;\lbd_n\geq\lbd_{n+1}\;$. 
Let $\;N\;$ be the largest integer for which $\;\sum_{n=1}^N\lbd_n\geq 0\;$.
Then the Lyapunov dimension [28] is
$$ D_L \equiv N + \frac{1}{|\lbd_{N+1}|}\sum_{n=1}^N\lbd_n . $$
If we denote by $\;\lbd_n^+\;$ positive Lyapunov exponents, then the 
metric entropy is $\;h\equiv\sum_n\lbd_n^+\;$. These definitions can 
be transformed to the language of multipliers as follows.

Let the local multipliers (29), at each fixed $\;t\;$, be ordered so that 
$$ |\mu_n(t)|\geq |\mu_{n+1}(t)| . $$
And let $\;N=N(t)\;$ be the largest integer for which
$$ \prod_{n=1}^{N(t)}|\mu_n(t)| \geq 1 . $$
Then we define the {\it local Lyapunov dimension} as
\be
D_L(t) \equiv N(t) +
\frac{\ln\prod_{n=1}^{N(t)}|\mu_n(t)|}{|\ln|\mu_{N+1}(t)||}.
\ee
The asymptotic Lyapunov dimension is
$$ D_L\equiv\lim_{t\ra\infty}D_L(t) , $$
if this limit exists. Similarly, the {\it local metric entropy} is
$$ h(t)=\frac{1}{t}\ln\prod_n|\mu_n^+(t)| \qquad (|\mu_n^+(t)| > 1) . $$

Concluding this section, it is worth making several remarks. First, the 
method of local multipliers seems to be a natural and convenient tool 
for analysing the local stability of motion. The appearance of 
local instability, in the case of quantum--mechanical models, may be 
connected with the quantum chaos [29] which is a temporal phenomenon. 
Note, however, that there are models of quantum systems [30,31] exhibiting 
the same asymptotic chaos as that in the models of classical mechanics.

Second, instead of emphasizing the time dependence, one could separate out 
one of the space coordinate considering the stability properties along 
the chosen space direction [32,33]. Then we could introduce the local, 
with respect to the separated variable, multipliers. Also, one could 
consider the motion treating a coupling parameter as a variable [34,35]. 
Then the stability properties with respect to the change of the 
coupling parameter could be analysed.

Third, we could be interested in the stability of stochastic processes 
against the variation, not of initial conditions, but of boundary conditions 
or of model parameters. Then a similar scheme of analysing the local 
stability could be developed.

\section{Examples of random processes}

The method described in the previous section will be illustrated here be 
several simple examples of random processes. We shall consider two types 
of the probability distribution $\;p(\xi)\;$ for the stochastic variable
$\;\xi\;$: the {\it uniform distribution}
$$ p(\xi) =\frac{1}{2\xi_0}\left [ \Theta(\xi + \xi_0) -
\Theta(\xi - \xi_0)\right ] , $$
in which $\;\Theta(\xi)\;$ is the unit--step function; and the {\it Gaussian
distribution}
$$ p_\ga(\xi) =\frac{1}{\sqrt{2\pi}\ga}\exp\left \{ -\frac{1}{2}\left (
\frac{\xi}{\ga}\right )^2\right \} . $$ 
The average value of $\;\xi\;$ in both cases is zero, and the average of $\;\xi^2\;$, respectively, is 
$$ \int_{-\infty}^{+\infty}\xi^2p(\xi)d\xi =\frac{1}{3}\xi_0^2 , \qquad 
\int_{-\infty}^{+\infty}\xi^2p_\ga(\xi)d\xi =\ga^2 . $$

\subsection{Random diffusion equation}

Consider the diffusion equation
\be
\frac{\prt y}{\prt t} =(D +\xi)\frac{\prt^2y}{\prt x^2} ,
\ee
in which $\;x\in[0,L],\; t\geq 0\;$, the diffusion coefficient $\;D > 0\;$, 
and $\;\xi\;$ describes random fluctuations of the diffusion coefficient. 
Such a situation can occur, for example, in a diffusion process through 
a nonhomogeneous medium consisting of randomly distributed regions 
with different diffusion coefficients. Eq.(31) is supplemented by an 
initial condition
$$ y(x,\xi,0) =f(x) $$
and boundary conditions
$$ y(0,\xi,t) =c_1, \qquad y(L,\xi,t) = c_2 . $$

Comparing (1) with (31), we see that the velocity field is
\be
v(x,y,\xi,t)=(D+\xi)\frac{\prt^2}{\prt x^2}y(x,\xi,t) .
\ee
For the Lyapunov matrix (10) we get
\be
L(x,x',\xi) = (D+\xi)\frac{\prt^2}{\prt x^2}\dlt(x -x') .
\ee
The equation (9) for the multiplier matrix is
\be
\frac{\prt M}{\prt t} = (D+\xi)\frac{\prt^2M}{\prt x^2} ,
\ee
where $\;M=M(x,x',\xi,t)\;$. Variating initial conditions and boundary
conditions gives
$$ M(x,x',\xi,0) =\dlt(x-x') , $$
$$ M(0,x',\xi,t) = M(L,x',\xi,t) = 0 . $$

Since the Lyapunov matrix (33) is stationary, it has common eigenfunctions 
with the multiplier matrix. These eigenfunctions, satisfying the 
boundary conditions
$$ \vp_n(0) = \vp_n(L) = 0 $$
and the normalization
$$ \int_{0}^{L}|\vp_n(x)|^2dx = 1 , $$
are 
$$ \vp_n(x) =
\sqrt{\frac{2}{L}}\sin k_nx \qquad (k_n\equiv\frac{\pi n}{L}) , $$
where $\;n=1,2,\ldots\;$. The basis $\;\{\vp_n(x)\}\;$ is stochastically
invariant, that is, does not depend on the random variable $\;\xi\;$.

The eigenvalues of (33) are
\be
\lbd_n(\xi) = -(D+\xi)k_n^2 .
\ee
Therefore, the local multipliers (17), with $\;t_0=0\;$, become
\be
\mu_n(\xi,t) =\exp\{-(D+\xi)k_n^2t\} .
\ee

The same answer can be obtained by solving the equation (34) yielding 
the multiplier matrix
\be
M(x,x',\xi,t) =\sum_{n=1}^\infty\mu_n(\xi,t)\vp_n(x)\vp_n(x') .
\ee
The eigenvalues of (37) are, evidently, given by (36).

The multiplier matrix (37) can also be found from the direct variation of 
the solution
$$ y(x,\xi,t) =\sum_{n=1}^\infty b_n\mu_n(\xi,t)\vp_n(x) + y_\infty(x) $$
to eq.(31), where
$$ b_n =\frac{1}{L}\int_0^L\left [ f(x) - y_\infty(x)\right ] \vp_n(x)dx , $$
$$ y_\infty(x) = c_1 + \frac{c_2-c_1}{L}x . $$

Now, let us find the averaged multiplier (25). For the case of the uniform
noise, the latter is
\be
\mu_n(t) =\frac{\sinh(\xi_0k_n^2t)}{\xi_0k_n^2t}\exp\left ( -Dk_n^2t\right ) .
\ee
The asymptotic behaviour of (38) at small and large times is
$$ \mu_n(t) \simeq 1 -Dk_n^2t \qquad (t\ra 0) , $$
\be
\mu_n(t)\simeq \frac{\exp\{(\xi_0 -D)k_n^2t\}}{2\xi_0k_n^2t} \qquad 
(t\ra\infty) .
\ee

According to (27) and (28), the behaviour of (38) defines the local stability 
at time $\;t\;$ of an $\;n\;$--component. Thus, the $\;n\;$--component for
$\;\xi_0 < D\;$ is uniformly stable for all $\;t > 0\;$. As $\;t\ra\infty\;$, 
it is exponentially stable. Since $\;\sup_n|\mu_n(t)| < 1\;$ for all 
$\;n\geq 1\;$, the process is uniformly stable for all $\;t > 0\;$.

When $\;\xi_0 =D\;$, the $\;n\;$--component is  also uniformly stable for 
all $\;t >0\;$. But as $\;t\ra\infty\;$, it displays now, not exponential, 
but power--law stability,
$$ \mu_n(t)\simeq (2\xi_0k_n^2t)^{-1} \qquad (t\ra\infty) . $$
The process is uniformly stable for $\;t>0\;$.

If $\;\xi_0 >D\;$, then the $\;n\;$--component is locally stable on the 
interval $\;(0,t_n)\;$, where $\;t_n\;$ is given by the condition
$\;|\mu_n(t_n)|=1\;$. After this, it becomes unstable for all $\;t > t_n\;$.
Because of the limit $\;\lim_{n\ra\infty}|\mu_n(t)|=\infty\;$, the process 
is uniformly unstable for any $\;t > 0\;$.

In the case of the noise with Gaussian distribution, the averaged multiplier
(25) becomes
\be
\mu_n(t) =\exp\left \{ \frac{1}{2}\left (\ga k_n^2t\right )^2 - 
Dk_n^2t\right \} .
\ee
In this case, the $\;n\;$--component for any $\;\ga>0\;$ is locally stable till 
$\;t_n=2D/(\ga k_n)^2\;$, when $\;|\mu_n(t_n)|=1\;$ and it is uniformly
unstable for $\;t > t_n\;$. As far as $\;\sup_n|\mu_n(t)|=\infty\;$, the 
process is uniformly unstable for all $\;t > 0\;$.

\subsection{Random wave equation}

In the wave equation
\be
\frac{\prt^2u}{\prt t^2} = (c+\xi)^2\frac{\prt^2u}{\prt x^2} ,
\ee
where $\;x\in[0,L],\; t\geq 0\;$, the inclusion of the random variable 
$\;\xi\;$ models fluctuations of the sound, or light, velocity $\;c\;$ 
in a randomly inhomogeneous medium. Write the initial conditions to (41),
$$ u(x,\xi,0) = f_1(x), \qquad u_t(x,\xi,0) = f_2(x) , $$
where $\;u_t\;$ means the derivative with respect to $\;t\;$. Take the 
boundary conditions as
$$ u(0,\xi,t) = 0, \qquad u(L,\xi,t) = 0. $$

To reduce (41) to the standard form (1), define
$$ y_1 \equiv u, \qquad y_2 \equiv \frac{\prt u}{\prt t} . $$
Then (41) is equivalent to the system
\be
\frac{\prt y_1}{\prt t} =y_2, \qquad 
\frac{\prt y_2}{\prt t} = (c +\xi)^2\frac{\prt^2y_1}{\prt x^2} .
\ee
The corresponding initial and boundary conditions are 
$$ y_1(x,\xi,0) = f_1(x) , \qquad y_2(x,\xi,0) = f_2(x) , $$
$$ y_1(0,\xi,t) = 0 , \qquad y_1(L,\xi,t) = 0 . $$
For the velocity field we have
$$ v_1(x,y,\xi,t) = y_2(x,\xi,t) , $$
\be
v_2(x,y,\xi,t) = (c +\xi)^2\frac{\prt^2}{\prt x^2} y_1(x,\xi,t) .
\ee
Thence, for the Lyapunov matrix (10) we get
\begin{eqnarray}
L(x,x',\xi)=\left [ \begin{array}{cc}
0 & 1 \\
(c+\xi)^2\frac{\prt^2}{\prt x^2} & 0 \end{array}\right ] \dlt (x-x') .
\end{eqnarray} 
Variating (41), we obtain the equations
$$ \frac{\prt M_{11}}{\prt t} = M_{21} , \qquad 
\frac{\prt M_{22}}{\prt t} = (c+\xi)^2\frac{\prt^2M_{12}}{\prt x^2} , $$
\be
\frac{\prt M_{12}}{\prt t} = M_{22} , \qquad 
\frac{\prt M_{21}}{\prt t} = (c+\xi)^2\frac{\prt^2M_{11}}{\prt x^2} 
\ee
for the multiplier matrix $\;M_{ij} =M_{ij}(x,x',\xi,t)\;$ with the initial 
and boundary conditions
$$ M_{ij}(x,x',\xi,0) =\dlt_{ij}\dlt(x-x') , $$
$$ M_{ij}(0,x',\xi,t) = 0, \qquad M_{ij}(L,x',\xi,t) = 0 . $$

Eigenfunctions of the Lyapunov matrix (44), satisfying the boundary  
and normalization conditions
$$ \vp_{ni}(0) = 0 , \qquad \vp_{ni}(L) = 0 , $$
$$ \int_0^L |\vp_{ni}(x)|^2dx = 1 \qquad (i=1,2) $$
are
\begin{eqnarray}
\vp_{n1}(x) =\sqrt{\frac{2}{L}}\frac{\sin k_nx}{\sqrt{1+\om_n^2}}
\left [ \begin{array}{c}
i \\
\om_n \end{array}\right ] , \qquad \vp_{n2}(x) =\vp_{n1}^*(x) ,
\nonumber
\end{eqnarray}
where
$$ \om_n =\om_n(\xi) = (c +\xi)k_n , \qquad k_n\equiv\frac{\pi n}{L} \quad
(n=1,2,\ldots ) . $$
Note that these eigenfunctions are not orthogonal to each other, since
$$ \int_0^L\vp_{n1}^+(x)\vp_{n2}(x)dx = \frac{\om_n^2-1}{\om_n^2+1} . $$
The eigenvalues of (44) are
\be
\lbd_{n1}(\xi) = i\om_n(\xi) , \qquad \lbd_{n2}(\xi) = -i\om_n(\xi) .
\ee
Due to the stationarity of the Lyapunov matrix (44), the multiplier matrix 
has the same eigenfunctions with the eigenvalues
\be
\mu_{n1}(\xi,t) =\exp\{ i\om_n(\xi)t\} , \qquad \mu_{n2}(\xi,t) =\mu_{n1}^*(\xi,t) .
\ee

Equations in (45) can be solved explicitly giving
\begin{eqnarray}
M(x,x',\xi,t) =2\sum_{n=1}^\infty\sin(k_nx)\sin(k_nx')\left [
\begin{array}{cc}
\cos\om_nt  & \frac{\sin\om_nt}{\om_n} \\
\\
-\om_n\sin\om_nt  &  \cos\om_nt 
\end{array} \right ] .
\end{eqnarray}
It can be checked that (48) has, really, the eigenfunctions $\;\vp_{ni}(x)\;$
and the eigenvalues (47). The multiplier matrix (48) could be also obtained 
from the direct variation, with respect to initial functions $\;f_1(x)\;$ and $\;f_2(x)\;$, of the solution
$$ u(x,\xi,t) =\sum_{n=1}^\infty \left [ A_n\cos\om_n(\xi)t + 
B_n\sin\om_n(\xi)t \right ]\sin k_nx , $$
in which 
$$ A_n =\frac{2}{L}\int_0^L f_1(x)\sin(k_nx)dx , $$
$$ B_n =\frac{2}{\om_n(\xi)L}\int_0^L f_2(x)\sin(k_nx)dx . $$

In the considered case, the eigenfunctions $\;\vp_{ni}(x)\;$ depend 
on $\;\xi\;$ through $\;\om_n(\xi)\;$. Therefore, eq.(26) is not valid. 
However, if the noise is weak, that is if $\;\xi_0\ll c\;$ 
or $\;\ga\ll c\;$, then, we can resort to perturbation theory with 
the zero--order basis $\;\{\vp_{ni}(x)\}\;$ in which $\;\om_n =\om_n(0)\;$. 
Then the spectrum $\;\om_n(\xi)\;$ is the first--order approximation.
Respectively, the averaged multiplier (25) can be defined as a first--order
approximation.

Averaging (47) with the uniform distribution, we get
$$ \mu_{n1}(t) = \frac{\sin(\xi_0k_nt)}{\xi_0k_nt}\exp (ick_nt) , $$
\be
\mu_{n2}(t) = \mu_{n1}^*(t) .
\ee
Since $\;|\mu_{n1}(t)|<1\;$ for all $\;n\geq 1\;$ and $\;t > 0\;$, the 
process is uniformly stable, its asymptotic stability being of power--law 
type.

For the Gaussian noise we find
$$ \mu_{n1}(t) =\exp\left \{ ick_nt - \frac{1}{2}\left ( \ga k_nt\right )^2
\right \} , $$
\be
\mu_{n2}(t)=\mu_{n1}^*(t) .
\ee
The process is uniformly stable for all $\;t>0\;$, and as $\;t\ra\infty\;$, 
it is exponentially stable.

Note that perturbation theory can be employed for defining the spectrum of 
the Lyapunov matrix. But after averaging over stochastic fields, the 
local multipliers (49) or (50) cannot be expanded in powers of $\;\xi_0\;$ 
or $\;\ga\;$ because now these parameters enter being factored by the 
time $\;t\;$.

\subsection{Random Schr\"odinger equation}

Take the  time--dependent Schr\"odinger equation
\be
i\frac{\prt\psi}{\prt t} = -\frac{1}{2m}\frac{\prt^2\psi}{\prt x^2} + 
(U +\xi)\psi
\ee
for a complex wave function $\;\psi=\psi(x,\xi,t)\;$, where $\;x\in(-\infty,+\infty),\; t\geq 0;\; m\;$ and $\;U\;$ are real constants, 
and the random variable $\;\xi\;$ imitates random fluctuations of 
the potential $\;U\;$. An initial conditions to (51) is
$$ \psi(x,\xi,0) = f(x) . $$

A special case of the initial condition will be considered here, when 
the initial function is periodic,
\be
f(x+L) = f(x) .
\ee
We opt for a periodic initial condition in order to compare the influence 
of random fields on the Schr\"odinger equation with the influence 
of nonlinearity. When periodic boundary conditions are enforced for the 
cubic Schr\"odinger equation, then its periodic solutions are well known to 
be subject to modulational long--wavelength instability [36,37]. The Cauchy
problem for the linear Schr\"odinger equation with periodic initial data, 
but without random fields, has also been intensively studied [38,39].
The behaviour of the solution, being closely related to incomplete Gaussian 
sums [39,40], has been found to look so chaotically as it would display 
the property of quantum chaos [41].

In the case of eq.(51), for the Lyapunov matrix we have
\be
L(x,x',\xi) =\left [ \frac{i}{2m}\frac{\prt^2}{\prt x^2} - i(U+\xi)\right ]
\dlt (x -x') .
\ee
Its eigenfunctions, satisfying the periodicity and normalization conditions
$$ \vp_n(x+L) =\vp_n(x) , \qquad \int_0^L|\vp_n(x)|^2dx = 1 , $$
are the plane waves
\be
\vp_n(x) =\frac{1}{\sqrt{L}}e^{-ik_nx} , \qquad k_n \equiv\frac{2\pi}{L}n ,
\ee
where $\;n\;$ is any integer. The eigenvalues of (53) are
\be
\lbd_n(\xi) = -i\left ( \frac{k_n^2}{2m} + U + \xi\right ) .
\ee
The Lyapunov matrix (53) is stationary, therefore the local multiplier is
\be
\mu_n(\xi,t) =\exp\left \{ -i\left ( \frac{k_n^2}{2m} + U + \xi\right )t
\right \} .
\ee

The multiplier matrix satisfies the equation
\be
\frac{\prt M}{\prt t} = i\frac{\prt^2M}{\prt x^2} - i(U +\xi)M
\ee
with the initial and periodicity conditions
$$ M(x,x',\xi,0) = \dlt (x-x') , $$
$$ M(x+L,x',\xi,t) = M(x,x',\xi,t) . $$
The solution to (57) is
\be
M(x,x',\xi,t) =\sum_{n=-\infty}^{+\infty}\mu_n(\xi,t)\vp_n(x)\vp_n^*(x') ,
\ee
from where it is evident that (58) has eigenfunctions (54) and eigenvalues 
(56). The same matrix (58) could be obtained by a direct variation of 
the solution
$$ \psi(x,\xi,t) =\sum_{n=-\infty}^{+\infty} c_n\mu_n(\xi,t)\vp_n(x) $$
to eq.(51), where
$$ c_n =\int_0^L f(x)\vp_n^*(x)dx . $$

The eigenfunctions (54) are stochastically invariant, which permits to 
define the averaged local multiplier (25). For the uniform noise one gets
\be
\mu_n(t) =\frac{\sin\xi_0t}{\xi_0t}
\exp\left \{ -i\left ( \frac{k_n^2}{2m} + U\right )t\right \} .
\ee
The process is uniformly stable for all $\;t>0\;$ with a power--law 
asymptotic stability.

For the Gaussian noise, the local multipliers are
\be
\mu_n(t) =\exp\left\{ -i\left (\frac{k_n^2}{2m} + U\right ) t -
\frac{1}{2}\left (\ga t\right )^2\right \} .
\ee
So, the process is uniformly stable for any $\;t>0\;$ with the exponential
asymptotic stability.

If random fields are absent, then $\;|\mu_n(0,t)|=1\;$, that is the 
motion is neutral. The solution $\;\psi(x,0,t)\;$ is quasiperiodic with 
a countable number of frequencies
$$ \om_n =\frac{1}{n}\left ( \frac{k_n^2}{2m} + U\right ) \qquad (n\neq 0) . $$
As is known, the behaviour of many--frequency quasiperiodic solutions can 
be quite complicated, often reminding chaotic one. But, of course, there 
is no chaos here. The neutral process, after the inclusion of random
fluctuations, becomes stable.

\section{Stability of quasiopen systems}

From the simple models of the previous section we now pass to the 
consideration of  a realistic statistical system. Take the the standard
Hamiltonian
$$ H =
\int\psi^\dgr(\sr)\left ( -\frac{\nabla^2}{2m} -\mu\right ) \psi(\sr)d\sr + $$
\be
+\frac{1}{2}\int\psi^\dgr(\sr)\psi^\dgr(\sr')\Phi(\sr -\sr')\psi(\sr')\psi(\sr)
d\sr d\sr'
\ee
of spinless particles with a chemical potential $\;\mu\;$, a symmetric
interaction potential $\;\Phi(\sr)=\Phi(-\sr)\;$ and with field operators
$\;\psi(\sr) \equiv\psi(\sr,t)\;$. Consider the density matrix
\be
\rho(\sr,\sr',t) =\lgl\psi^\dgr(\sr',t)\psi(\sr,t)\rgl ,
\ee
in which the brackets $\;\lgl\ldots\rgl\;$ mean the statistical averaging 
with a statistical operator $\;\hat\rho(0)\;$. The diagonal element of (62) 
is the density of particles
\be
\rho(\sr,t) \equiv\rho(\sr,\sr,t) \geq 0 .
\ee
The average density of particles in a system of volume $\;V\;$ is
\be
\frac{1}{V}\int_V\rho(\sr,t)d\sr =\rho .
\ee
In the thermodynamic limit, $\;V\ra\infty\;$, the right--hand side of (64)
remains constant. Another property of the density matrix (62) following 
from its definition is
\be
\rho^*(\sr,\sr',t) =\rho(\sr',\sr,t) .
\ee
Differentiating (62) with respect to time and invoking the Heisenberg 
equations for the field operators, one gets the evolution equation 
relating (62) with the two--particle density matrix
\be
\rho_2(\sr_1,\sr_2,\sr_1',\sr_2',t) =
\lgl\psi^\dgr(\sr_1')\psi(\sr_2')\psi(\sr_2)\psi(\sr_1)\rgl ,
\ee
in which all field operators contain the same time variable $\;t\;$, that is,
$\;\psi(\sr_i)\equiv\psi(\sr_i,t)\;$.

Suppose that the considered system is {\it randomly open} in the sense 
that it is subject to the action of random forces from surrounding whose
influence can be taken into account by supplementing the evolution equation 
with a random field. Thus, the evolution equation for the density 
matrix (62) takes the form
$$ i\frac{\prt}{\prt t}\rho(\sr_1,\sr_2,\xi,t) =
\left [ -\frac{1}{2m}\left ( \nabla_1^2 - \nabla_2^2\right ) + \xi\right ]
\rho(\sr_1,\sr_2,\xi,t) + $$
\be
+\int\left [\Phi(\sr_1 -\sr_3) - \Phi(\sr_2 -\sr_3)\right ]
\rho_2(\sr_1,\sr_3,\sr_2,\sr_3,\xi,t)d\sr_3 ,
\ee
where the random variable $\;\xi\;$ can, in general, be complex. If we 
put $\;\xi=0\;$ in (67), then we return to the usual evolution equation for 
an isolated system.

Emphasize the difference between a random isolated system and a 
random open system. In the former case, one should add random 
fields into the Hamiltonian (61), so such fields are to be Hermitian. 
In the latter case, random fields are to be inserted into the 
evolution equation, and they are not necessarily self--adjoint.

To make the evolution equation (67) closed, we need to resort to 
an approximation for the two--particle density matrix (66). Let us use
the Hartree--Fock approximation
\be
\rho_2(\sr_1,\sr_2,\sr_1',\sr_2',t)=\rho(\sr_1,\sr_1',t)\rho(\sr_2,\sr_2',t)\pm
\rho(\sr_1,\sr_2',t)\rho(\sr_2,\sr_1',t) .
\ee
If the interaction potential $\;\Phi(\sr)\;$ is strongly singular, then 
the Hartree--Fock approximation leads to divergences. In that case, one must 
use the correlated Hartree--Fock approximation [42], in which the 
interaction potential is smoothed by a correlation function. Keeping in 
mind these both possibilities, we imply in what follows that 
$\;\Phi(\sr)\;$ is integrable. The evolution equation (67) becomes
$$ \left [ i\frac{\prt}{\prt t} +\frac{1}{2m}\left ( \nabla_1^2 - 
\nabla_2^2\right ) -\xi\right ]\rho(\sr_1,\sr_2,\xi,t) = $$
$$ = \int\left [\Phi(\sr_1 -\sr_3) -\Phi(\sr_2 -\sr_3)\right ]
\left [\rho(\sr_1,\sr_2,\xi,t)\rho(\sr_3,\sr_3,\xi,t) \pm\right. $$
\be
\left. \pm \rho(\sr_1,\sr_3,\xi,t)\rho(\sr_3,\sr_2,\xi,t)\right ]d\sr_3 .
\ee

The solution to eq.(69) has to satisfy conditions (63)--(65) and an 
initial condition
\be
\rho(\sr_1,\sr_2,\xi,0) = f(\sr_1,\sr_2) .
\ee

To analyse the stability of processes described by eq.(69), we need to 
consider the multiplier matrix
\be
M(\sr_1,\sr_2,\sr_1',\sr_2',\xi,t) =
\frac{\dlt\rho(\sr_1,\sr_2,\xi,t)}{\dlt\rho(\sr_1',\sr_2',\xi,t)}
\ee
with an initial condition
\be
M(\sr_1,\sr_2,\sr_1',\sr_2',\xi,0) = \dlt(\sr_1 -\sr_1')\dlt(\sr_2 -\sr_2') .
\ee
For the multiplier matrix (71) we may write the evolution equation (9) with 
the Lyapunov matrix
\be
L(\sr_1,\sr_2,\sr_1',\sr_2',\rho,\xi,t) =
\frac{\dlt v(\sr_1,\sr_2,\rho,\xi,t)}{\dlt \rho(\sr_1',\sr_2',\xi,t)} ,
\ee
where the velocity field is
$$ v(\sr_1,\sr_2,\rho,\xi,t) = \left [ \frac{i}{2m}\left ( \nabla_1^2 - 
\nabla_2^2\right ) - i\xi\right ]\rho(\sr_1,\sr_2,\xi,t) - $$
$$  - i\int\left [\Phi(\sr_1 -\sr_3) -\Phi(\sr_2 -\sr_3)\right ]
\left [\rho(\sr_1,\sr_2,\xi,t)\rho(\sr_3,\sr_3,\xi,t) \pm\right. $$
\be
\left. \pm \rho(\sr_1,\sr_3,\xi,t)\rho(\sr_3,\sr_2,\xi,t)\right ]d\sr_3 .
\ee
From (73) and (74) we find
$$ L(\sr_1,\sr_2,\sr_1',\sr_2',\rho,\xi,t) = \left [ \frac{i}{2m}\left ( \nabla_1^2 - \nabla_2^2\right ) - i\xi\right ]
\dlt(\sr_1 -\sr_1')\dlt(\sr_2 -\sr_2') - $$
$$ -i\dlt(\sr_1 -\sr_1')\dlt(\sr_2 -\sr_2')
\int\left [\Phi(\sr_1 -\sr_3) -\Phi(\sr_2 -\sr_3)\right ] \rho(\sr_3,\sr_3,\xi,t)d\sr_3 - $$
$$ -i\left [\Phi(\sr_1 -\sr_1') -\Phi(\sr_2 -\sr_1')\right ] \rho(\sr_1,\sr_2,\xi,t)\dlt(\sr_1' -\sr_2')\mp $$
\be
\mp i\left [\Phi(\sr_1 -\sr_1') -\Phi(\sr_2 -\sr_1')\right ] \rho(\sr_1,\sr_1',\xi,t)\dlt(\sr_2 -\sr_2')\mp 
\ee
$$ \mp i\left [\Phi(\sr_1 -\sr_2') -\Phi(\sr_2 -\sr_2')\right ] \rho(\sr_2',\sr_2,\xi,t)\dlt(\sr_1 -\sr_1') . $$

Consider the case when the random field $\;\xi\;$ is weak. More 
precisely, this means that the average of $\;|\xi|^2\;$ is much 
less than the average energy of the system. Of great importance is the 
limiting situation when the average of $\;|\xi|^2\;$ tends to $\;+0\;$, 
that is, when the random field is arbitrary small although finite. 
A statistical system with the evolution equation (67) subject to the 
action of an asymptotically small random filed will be called 
{\it randomly quasiopen}.

If in eq.(69) we put $\;\xi\equiv 0\;$, then it is easy to check that 
the density matrix
\be
\rho(\sr_1,\sr_2,0,t) =\rho_0(\sr_1 -\sr_2) ,
\ee
where $\;\rho_0(\sr)\;$ is an arbitrary function satisfying just two conditions
\be
\rho_0^*(\sr) =\rho_0(-\sr) , \qquad \rho_0(0)=\rho > 0 ,
\ee
is a stationary solution.Then the density of particles
$$ \rho(\sr,\sr,0,t) =\rho_0(0) =\rho $$
is uniform in real space. It is worth stressing that there is infinite 
number, actually, a functional continuum, of stationary solutions (76), 
for which conditions (77) are fulfilled.

A common convention, concluded from the existence of stationary 
solutions (76), is that an isolated statistical system with any 
initial condition will finally, as $\;t\ra\infty\;$, tend to a stationary 
state called the state of absolute equilibrium. The fact that such a state 
is infinitely degenerate, in the sense that there exist infinitely many 
density matrices with the same diagonal part but different nondiagonal parts, 
is interpreted as follows. All these density matrices are treated as 
statistically equivalent provided that they define the same set of 
statistical averages for local observables. As is obvious, the convention 
about the existence of absolute equilibrium is based on the assumption of 
its asymptotic stability. 

Using the method of multipliers we can check the stability of a stationary
solution given by (76). The Lyapunov matrix with the solution (76) 
is stationary, i.e. according to the notation (13),
$$ L(\xi) =\left [ L(\sr_1,\sr_2,\sr_1',\sr_2',\rho_0,\xi)\right ] $$
does not depend on time. When there are no random fields, that 
is $\;\xi\equiv 0\;$, then the  eigenfunctions of $\;L(0)\;$ are arbitrary 
functions of the type $\;\vp(\sr_1 -\sr_2)\;$ with the eigenvalue
$\;\lbd(0)=0\;$. The corresponding motion is neutral, as it should be 
expected for a mean--field approximation. In the presence of weak 
random fields, the eigenvalue problem for $\;L(\xi)\;$ can be solved 
using perturbation theory. Then in the first--order approximation
$$ \lbd(\xi) =-i\xi . $$
And for the local multiplier we have
\be
\mu(\xi,t) =\exp(-i\xi t) .
\ee

Consider the averaged local multiplier (25) assuming that the random
variable $\;\xi\;$ pertains to the complex plane. For the uniform
distribution, let $\;Re\xi\in[-\xi_1,+\xi_1]\;$ and
$\;Im\xi\in[-\xi_2,+\xi_2]\;$. Then
\be
\mu(t)=\frac{\sin(\xi_1t)\sinh(\xi_2t)}{\xi_1\xi_2t^2} .
\ee
The asymptotic behaviour of (80) is
$$ \mu(t)\simeq 1 -\frac{1}{6}\left ( \xi_1^2-\xi_2^2\right ) t^2 \qquad 
(t\ra 0) , $$
$$ \mu(t) \simeq \frac{\sin(\xi_1t)}{2\xi_1\xi_2t^2}e^{\xi_2t} \qquad
(t\ra\infty) . $$

If $\;\xi_1\geq\xi_2\;$, then the motion is locally stable at small time. 
With increasing time the stability is lost, but it is recovered around 
the recurrence time $\;t_{rec}=\frac{\pi}{\xi_1}n\;(n=1,2,\ldots)\;$, 
where $\;\mu(t_{rec})=0.\;$. So, there occurs a peculiar {\it stability echo}.
When $\;\xi_1=\xi_2\;$, then the first region of local stability is inside 
the interval $\;0<\xi_1t<3.8\;$; the second, inside the interval
$\;6.1<\xi_1t<6.4\;$, and so on. There is no asymptotic stability, but 
the regions of local stability and instability change one another. Such a 
kind of behaviour is usually called {\it intermittent}. If $\;\xi_1<\xi_2\;$,
then the motion at small time is unstable, but with increasing time it is 
again intermittent, when instability regions are interrupted by the 
stability echo. Here the intermittency happens with respect to time, but it 
may occur in space as well [43].

If the noise is Gaussian, we denote by $\;\ga_1\;$ the dispersion 
of $\;Re\xi\;$; and by $\;\ga_2\;$, the dispersion of $\;Im\xi\;$. 
Then the averaged local multiplier is
\be
\mu(t) =\exp\left\{-\frac{1}{2}\left (\ga_1^2 -\ga_2^2\right ) t^2\right \} .
\ee
For $\;\ga_1>\ga_2\;$, the motion is uniformly stable for all $\;t>0\;$. 
If $\;\ga_1 =\ga_2\;$, the motion is neutral. And it is uniformly unstable 
for any $\;t>0\;$, if $\;\ga_1<\ga_2\;$.

In this way, for both uniform as well as Gaussian noise there exist such 
small random fields that make the stationary solutions for the 
density matrix unstable. In the presence of such fields, the solution 
to (69) will wander between infinite many of the stationary solutions, 
always remaining unstable and nonstationary. Therefore, the density of 
particles (63) will overlastingly depend on time fluctuating in real space.
Accepting that no realistic statistical system can be ideally isolated, 
but should be treated rather as randomly quasiopen, we come to 
the conclusion that such a system never reaches the state of absolute
equilibrium. A randomlyly quasiopen system can become not more 
than quasiequilibrium, with perpetually persisting mesoscopic 
fluctuations [44].

\section{Random structural stability}

An important question is how the qualitative behaviour of a dynamical system
without random fields changes when the latter are switched on. Define by
$\;\dlt=\{\dlt_i\}\;$ a set of parameters characterizing the distribution 
of random variables. For example, in the case of real random fields,
$\;\dlt=\xi_0\;$ for the uniform distribution, and $\;\dlt=\ga\;$ for 
the Gaussian distribution. In the case of complex random fields,
$\;\dlt=\{\xi_1,\xi_2\}\;$ for the uniform noise, and $\;\dlt=\{\ga_1,\ga_2\}\;$
for the Gaussian noise. A dynamical system without random fields 
corresponds to $\;\dlt=0\;$, which implies that all $\;\dlt_i=0\;$.

Define the {\it parametric neighborhood} of $\;\dlt=0\;$ as a manifold
$\;\{\dlt_i\in[0,\ep_i] \;|\ep_i>0\}\;$ of stochastic parameters 
$\;\dlt_i\;$ pertaining to arbitrary small finite intervals $\;[0,\ep_i]\;$ 
with $\;\ep_i\;$ independent of each other.

We shall say that the $\;n\;$--component is {\it random--structurally
stable} with respect to a given random field, if there exists a 
parametric neighborhood of $\;\dlt=0\;$ such that
$$ \lim_{t\ra\infty}\lim_{\dlt\ra 0}|\mu_n(t)|=
\lim_{\dlt\ra 0}\lim_{t\ra\infty}|\mu_n(t)| $$
for any sequence of $\;\dlt\;$ from this parametric neighborhood. The limit
$\;\dlt\ra 0\;$ means the all $\;\dlt_i\ra 0\;$. The commutativity of the 
limits in the above equality can be expressed shorter as
$$ \left [\lim_{t\ra\infty},\lim_{\dlt\ra 0}\right ]\left |\mu_n(t)\right | =
0 . $$
When this commutativity holds true for all $\;n\;$--components, we shall 
say that the {\it dynamical process is random--structurally stable}.

Let us illustrate the notion of this stability for the examples of 
section 3. Thus, for the random diffusion equation we have
$$ \lim_{t\ra\infty}\lim_{\dlt\ra 0}|\mu_n(t)| = 0 $$
for all $\;n=1,2,\ldots\;$. And the opposite order of the limits gives
\begin{eqnarray}
\lim_{\dlt\ra 0}\lim_{t\ra\infty}|\mu_n(t)| =\left \{
\begin{array}{cc}
0 & (uniform \; noise) \\
\nonumber
\infty & (Gaussian\; noise)
\end{array}\right.
\end{eqnarray}
again for any $\;n\geq 1\;$. Therefore, the diffusion process is 
random--structurally stable with respect to the uniform noise and
unstable with respect to the Gaussian noise.

For the random wave equation we get
$$ \lim_{t\ra\infty}\lim_{\dlt\ra 0}|\mu_{ni}(t)| = 1 $$
for all $\;n=1,2,\ldots\;$ and $\;i=1,2\;$. While
$$ \lim_{\dlt\ra 0}\lim_{t\ra\infty}|\mu_{ni}(t)| = 0 $$
for both uniform and Gaussian noise, and any $\;n\geq 1\;$ and 
$\;i=1,2\;$. Consequently, the wave process is random--structurally unstable 
with respect to the uniform as well as to the Gaussian noise.

Similarly to the previous case, for the random Schr\"odinger equation we 
find
$$ \lim_{t\ra\infty}\lim_{\dlt\ra 0}|\mu_n(t)| = 1 , $$
$$ \lim_{\dlt\ra 0}\lim_{t\ra\infty}|\mu_n(t)| = 0  $$
for all $\;n\in\Bbb{N}\;$ and for both uniform and Gaussian noise. Thus, 
the process is also random--structurally unstable with respect to these noises.

For the randomly quasiopen system of sec.4, we obtain
$$ \lim_{t\ra\infty}\lim_{\dlt\ra 0}|\mu(t)| = 1 . $$
But interchanging the limits makes the result undefined. For the uniform noise 
we get the intermittent behaviour, and for the Gaussian noise we can come 
to $\;-\infty,\; 1\;$ or $\;+\infty\;$ depending on the relation 
between $\;\ga_1\;$ and $\;\ga_2\;$. Thence, the dynamical process 
for a randomly quasiopen system is random--structurally unstable.

The notion of random--structural stability is analogous to that of structural
stability [28,45]. Neutral dynamical processes are known to be structurally 
unstable. As we have seen, such processes are also random--structurally
unstable.

\end{document}


{\large{\bf Appendix. Properties of multiplier matrices}}

\vspace{3mm}

In the stability analysis of random dynamical processes we have used 
the properties of multiplier matrices following from the conditions $\;1\;$ 
and $\;2\;$ of sec.2. Here we shall consider some additional properties of 
these matrices related to other conditions. For simplicity, we put 
$\;t_0=0\;$ dealing with $\;M(\xi,t)\equiv M(\xi,t,0)\;$. Then eq.(9) takes 
the form
$$ \frac{\prt}{\prt t}M(\xi,t) = L(y,\xi,t)M(\xi,t) , $$
\be
M(\xi,0) = 1. 
\ee

Suppose that the eigenvalue problem for the multiplier matrix has a solution 
with the eigenfunctions forming a complete orthonormal basis:
$$ M(\xi,t)\vp_n(\xi,t) =\mu_n(\xi,t)\vp_n(\xi,t) , $$
\be
\vp_m^+(\xi,t)\vp_n(\xi,t)=\dlt_{mn} ,
\ee
$$ \sum_{n}\vp_n(\xi,t)\vp_n^+(\xi,t) =\hat 1 . $$

\vspace{3mm}

{\bf Property 1.} If a multiplier matrix satisfying conditions (A.1) and 
(A.2) is Hermitian, then its eigenvalues are
$$ \mu_n(\xi,t) =\exp\left\{\int_0^t L_n(y,\xi,t')dt'\right\} $$
where
$$ L_n(y,\xi,t) \equiv \vp_n^+(\xi,t)L(y,\xi,t)\vp_n(\xi,t) . $$

\vspace{2mm}

{\it Proof.} Differentiating the normalization equality in (A.2), we get
$$ \frac{\prt\vp_m^+}{\prt t}\vp_n +\vp_m^+\frac{\prt\vp_n}{\prt t} = 0. $$
Using this, we find
$$ \frac{\prt}{\prt t}\left (\vp_m^+ M\vp_n\right ) =
\vp_m^+\frac{\prt M}{\prt t}\vp_n + (\mu_m^* - \mu_n)\vp_m^+
\frac{\prt\vp_n}{\prt t} . $$
Since, by assumption, the multiplier matrix is Hermitian, $\;M^+=M\;$, 
its eigenvalues are real, $\;\mu_n^*=\mu_n\;$. Because of this, we have 
the equality
$$ \vp_n^+\frac{\prt M}{\prt t}\vp_n = \frac{\prt\mu_n}{\prt t} . $$
The latter, together with (A.1) and (A.2), gives
$$ \frac{\prt\mu_n}{\prt t} =L_n\mu_n , \qquad \mu_n(\xi,0) = 1 , $$
from where we obtain Property 1.

Note that $\;L_n\;$ here is not necessarily an eigenvalue of the 
Lyapunov matrix, but just a matrix element of the latter with respect to 
the eigenfunctions of the multiplier matrix. These eigenfunctions of 
$\;M\;$ are not the eigenfunctions of $\;L\;$.

Another condition which yields an explicit relation between the multiplier 
and Lyapunov matrices is the commutation relation
\be
\left [ L(y,\xi,t),\int_0^t L(y,\xi,t')dt'\right ] = 0 ,
\ee
which sometimes called the Lappo--Danilevsky condition [26].

\vspace{3mm}

{\bf Property 2.}
Let the multiplier matrix enjoy conditions (A.1), (A.2), (A.3) and be 
Hermitian, then it possesses a common set of  eigenfunctions with the 
Lyapunov matrix, their eigenvalues being related as
$$ \mu_n(\xi,t)=\exp\left\{\int_0^t\lbd_n(\xi,t')dt'\right\} , $$
where $\;\lbd_n\;$ is an eigenvalue of the Lyapunov matrix,
$$ L(y,\xi,t)\vp_n(\xi,t) =\lbd_n(\xi,t)\vp_n(\xi,t) . $$

\vspace{2mm}

{\it Proof.} From (A.1) and (A.3) it follows that
$$ M(\xi,t) =\exp\left\{\int_0^t L(y,\xi,t')dt'\right\} , $$
so that $\;[M,L]=0\;$. Commuting matrices have a common set of 
eigenfunctions. Because of the self--adjointness of the multiplier 
matrix, Property 1 holds true. Since now $\;\vp_n\;$ are the eigenfunctions 
of the Lyapunov matrix, we have $\;L_n=\lbd_n\;$, which accomplishes the 
proof.

\vspace{2mm}

The properties considered can simplify the consideration. However, a 
principal difficulty remains, related to the fact that the eigenfunctions
$\;\vp_n(\xi,t)\;$ depend, generally, on the random variable $\;\xi\;$ 
as well as on time $\;t\;$.

\vspace{3mm}

{\bf Property 3.} Assume that the multiplier matrix satisfies 
condition (A.2), then the variation (11) yields
$$ \dlt c_n(\xi,t) =\mu_n(\xi,t)\dlt f_n(\xi,t) , $$
where
$$ c_n(\xi,t) =\vp_n^+(\xi,t)y(\xi,t) , $$
$$ f_n(\xi,t) =\vp_n^+(\xi,t) f . $$

\vspace{2mm}

{\it Proof.} The proof is straightforward.

\vspace{3mm}

The latter property shows that, because of the dependence of $\;f_n(\xi,t)\;$ 
on the random variable, one cannot, in general, define the simple 
relation (26) with the averaged local multiplier (25). This multiplier can 
be defined if the eigenfunctions $\;\vp_n(\xi,t)\;$ are stochastically
invariant, that is $\;\vp_n(\xi,t)=\vp_n(t)\;$, or if random fields 
are weak and the stochastic invariance of $\;\vp_n(t)\;$ is an acceptable
approximation. In such a case the equation from  Property 3 reduces to 
$$ \dlt c_n(t) =\mu_n(t)\dlt f_n(t) , $$
where $\;c_n(t)\;$ and $\;\mu_n(t)\;$ are the same as in (23) and (25) 
and $\;f_n(t) =\vp_n^+(t)f\;$.

However, in the obtained variation the multiplier $\;\mu_n(t)\;$ does not 
play the role of a quantity characterizing stability with respect to 
the variation of initial conditions, since $\;f_n(t)\;$ depends on time. 
Only if the eigenfunctions $\;\vp_n(t)\;$ are stationary, i.e.
$\;\vp_n(t)=\vp_n\;$, then we return to (26), and the local 
multiplier (25) becomes a genuine characteristic of local stability. 
This is why the condition of stationarity (condition 1 of sec.2) 
has been essential for the stability analysis.